\documentclass[fleqn,12pt]{article}

\usepackage{amsmath,amssymb}
\usepackage{amsfonts}
\usepackage{amsthm}
\usepackage{epsf}
\usepackage{graphicx,graphics}
\usepackage{fullpage}
\usepackage{setspace}
\usepackage[font=small,format=plain,label font=bf,up, textfont=it,up]{caption}
\usepackage[round]{natbib}
\usepackage[title]{appendix}
\usepackage{mwe}
\usepackage{caption}
\usepackage{subcaption}

\graphicspath{{/home/basu15/Dropbox/Marx-Economics/Manuscript/}}

\renewcommand{\epsilon}{\varepsilon}

\newtheorem{corollary}{Corollary}
\newtheorem{theorem}{Theorem}
\newtheorem{assumption}{Assumption}
\newtheorem{property}{Property}

\renewcommand{\epsilon}{\varepsilon}

\begin{document}
	
	\bibliographystyle{apalike}
	
	\title{Marx after Okishio: Falling Rate of Profit with Constant Rate of Exploitation
		\\
		\hspace*{1cm}
		\\
	}
	\author{Deepankar Basu\thanks{Department of Economics, University of Massachusetts Amherst,
			310 Crotty Hall, 412 N. Pleasant Street, Amherst MA 01002. Email: {\tt dbasu@econs.umass.edu}. We would like to thank Weikai Chen, Debarshi Das, Thomas R. Michl, Hyun Woong Park, and Naoki Yoshihara for comments on an earlier version of this paper. The usual disclaimers apply.} 
		\and
		Oscar Orellana\thanks{Departamento de Matem\'{a}tica, Universidad T\'{e}cnica Federico Santa Maria, Avenida Espa\~{n}a 1680, Valpara\'{i}so-Chile. Email: {\tt oscar.orellana@usm.cl}. This author acknowledges financial support provided by ANID Santiago-Chile under the Proyecto Fondecyt Regular n\'{u}mero 1181414 and Universidad T\'{e}cnica Federico Santa María, Valparaíso-Chile.}
		\\ \hspace*{1cm}}
	
	\date{\today}

	\maketitle

	\begin{abstract}
		\noindent
		Can cost-reducing technical change lead to a fall in the long run rate of profit if class struggle manages to keep the rate of exploitation constant? In a general circulating capital model, we derive sufficient conditions for cost-reducing technical change to both keep the rate of exploitation constant and lead to a fall in the equilibrium rate of profit. Further, if the real wage bundle is such that the maximum price-value ratio is larger than $1$ plus the rate of exploitation, then starting from any configuration of technology and real wage, we can always find a viable, CU-LS technical change that satisfies the sufficient conditions for the previous result. Taken together, these results vindicate Marx's claim in Volume III of \textit{Capital}, that if the rate of exploitation remains unchanged then viable, CU-LS technical change in capitalist economies can lead to a fall in the long run rate of profit. 
	\end{abstract}
	
	\doublespacing
	\section{Introduction}
	Whether the rate of profit has a tendency to decline with capitalist development was of considerable interest to classical political economy. Being the income stream of the capitalist class, profit is both the source and spur for capital accumulation. If there was a tendency for the rate of profit to continuously fall over time, this naturally pointed to some deep contradiction in the capitalism system. For, through this tendency, the system seemed to undermine itself \citep[Chapter~IV]{dobb_1945}.  
	
	Adam Smith had argued that capital accumulation and competition between capitalists would impart a tendency to the rate of profit to fall. David Ricardo, while disagreeing with Smith's explanation, nonetheless fell compelled to offer his own answer. Diminishing returns on land, argued Ricardo, was the ultimate source of the tendency for the rate of profit to fall. For, with capital accumulation, there is a rise in the demand for labor and therefore for food. This, in turn, necessitates the cultivation of inferior plots of land, thereby raising the price of labor and squeezing profits \citep[Chapter~IV, pp. 86--87]{dobb_1945}. Ricardo's argument shifted the locus of the declining tendency of the rate of profit to outside capitalism, to the possibilities or otherwise of technical progress in agricultural production. Marx brought the focus back to the internal dynamics of capitalism. 
	
	In developing his own argument about the law of the tendential fall in the rate of profit in Volume III of \textit{Capital}, Marx argued that technical progress in capitalist production, which brings about a rise in the organic composition of capital (the ratio of material and labor costs), would manifest itself as a `tendency' of the rate of profit to fall \citep{marx_3}. It is immaterial whether there is technical progress in agriculture. As long as the organic composition of capital has a tendency to rise at the aggregate level, technical progress, the very strength of capitalism, will undermine itself by imparting a declining trend to the rate of profit.
	
	Starting with \citet{okishio_1961}, a large literature has argued that Marx's argument is flawed. If capitalists adopt a new techniques of production only if they reduce the cost of production at existing prices, which seems reasonable, then the rate of profit will have a tendency to rise, rather than to fall. To be more precise, if capitalist producers choose to adopt a new technique of production only if it is cost-reducing at current prices and the real wage rate remains unchanged, then the long run rate of profit in the economy will rise \citep{okishio_1961, bowles_1981, roemer_1981, dietzenbacher_1989}. 
	
	
	Okishio's justly celebrated result rests on the assumption that the real wage rate does \textit{not} change. This is an extremely restrictive assumption, given that the analysis is about \textit{long run} prices and profit rates. There is no theoretical or empirical reason to believe that the real wage rate remains constant over the course of technical change, i.e. the adoption of a new technique of production by an innovating capitalist and its subsequent diffusion through the rest of the economy.\footnote{Even \citet{okishio_2000} admits that the assumption of a constant real wage rate is unrealistic. ``The assumption of a constant real wage rate implies either a non-monetary economy or the instantaneous adaptation of the money-wage rate to the prices of consumption goods. Both are unrealistic. A capitalistic economy is a monetary production economy. Labourers receive a money-wage. The money-wage rate and the prices of consumption goods change owing to competition in the consumption goods market and in the labour market. The assumption of a constant real wage rate cannot be maintained.'' \citep[pp.~493]{okishio_2000}.} In fact, technical change interacts with larger social and economic forces, including those relevant to labour market outcomes, and it is not inconceivable that the real wage rate can change - one way or the other - after technical change. 
	
	Taking a Marxian view of the matter suggests that the real wage rate is an outcome of class struggle, and it is unclear why class struggle would not be able to change the real wage rate over the course of technical change. At the least its seems plausible to argue that, since technical change increases labor productivity, workers will attempt to bargain for some part of the gain of technical change, especially in a context of labor constraints \citep[pp.~113--114]{dobb_1945}. Hence, it is eminently possible that the real wage rate will \textit{increase} with technical change, rather than remain unchanged in advanced capitalist economies marked by labor constraints. An important finding of the Marxist literature on technical change and distribution, one that is often not appreciated, is that Okishio's result will no longer hold if we allow the real wage to change over the course of technical change \citep{roemer_1981, foley_1986b, dietzenbacher_1989, laibman_1992, liang_2021}. 
	
	There have been two broad approaches to providing more structure to how the real wage rate might change over the course of technical change, i.e. discovery, adoption and diffusion of new techniques of production. The first approach has worked with a \textit{constant wage-profit ratio} as a plausible description of how the real wage rate might behave over the course of technical change. An analysis of the effect of technical change on the rate of profit when the profit-wage ratio remains constant was worked out in a $2$-commodity model in \citet{roemer_1977, roemer_1981}. Two important findings in \citet{roemer_1977} are that, first, we can only define sectoral profit-wage ratios, but not the aggregate profit-wage ratio, without reference to the scale of production, and second, that sectoral profit-wage ratios can remain constant only when the real wage varies across sectors, i.e. we need to assume non-competitive labour markets. The main result in \citet{roemer_1977, roemer_1981} is that the rate of profit falls (or remains unchanged) if there is cost-reducing capital-using labour-saving (CU-LS) technical change in the capital goods (consumer goods) sector. This result has been generalized to the case of an $n$-commodity model - without distinguishing between capital and consumer goods industries - in \citet{chen_2019}, which shows that when there is cost-reducing CU-LS technical change in any sector with sectoral profit-wage ratios remaining constant, the equilibrium profit rate falls.
	
	The second approach uses a \textit{constant rate of exploitation} as a description of how the real wage might vary over the course of technical change. The idea that the rate of exploitation might remain constant before and after technical change goes back to Marx \citep{marx_3}. His analysis of the law of the tendential fall worked with the often implicit assumption of a constant rate of exploitation. \citet{laibman_1982,laibman_1992} incorporated this assumption in a two-sector model and analysed the effect of technical change on the rate of profit. The main finding of \citet{laibman_1982} was that it is \textit{possible} for the rate of profit to fall after cost-reducing technical change if the rate of exploitation remains constant. While \citet{michl_1988} and \citet[Chapter~6]{basu_2021} presents similar results in a one sector model, \citet{liang_2021} has generalized Laibman's result to an $m$-sector two department model with fixed capital.
	
	This paper contributes to this literature by extending the results in \citet{laibman_1982}, \citet{michl_1988}, \citet[Chapter~6]{basu_2021} and \citet{liang_2021}. We extend the analysis of \citet{laibman_1982}, \citet{michl_1988} and \citet[Chapter~6]{basu_2021} to a general $n$-sector circulating capital model of a capitalist economy. Unlike \citet{laibman_1982}, we do not distinguish between capital and consumption goods. We extend the analysis in \citet{liang_2021} by allowing for a general change in the real wage bundle. Whereas \citet{liang_2021} only allows proportional changes in the vector of the real wage bundle, we allow for the real wage bundle to change in an arbitrary manner over the course of technical change. In this general setting, we demonstrate that under certain plausible conditions, the long run rate of profit can fall after viable technical change if the rate of exploitation remains constant (or even rises in a bounded manner). One advantage of using the constant rate of exploitation description of real wage behavior is that we do not need to assume non-competitive labor markets, as is needed in \citet{roemer_1981} and \citet{chen_2019}.
	
	The intuition for our result is straightforward. When a new technique of production becomes available in a sector, capitalists compare the cost of production associated with the new technique and the old technique using \textit{current} prices and wage rates. Capitalists do not know the direction in which class struggle will proceed and therefore do not take account of possible changes in the nominal or real wage rate - an outcome of class struggle - when arriving at their decision to adopt the new technique of production. Hence, if the technique reduces costs of production at current prices and wage rates, capitalists adopt the new technique of production.
	
	The course of class struggle can, under certain circumstances, lead to an increase in the real wage bundle in such a way that it not only becomes more expensive at current prices but also keeps the rate of exploitation unchanged. If technical change is of the capital-using labor-saving (CU-LS) type, the predominant form of technical change in capitalism \citep{foley_michl__tavani_2019}, then the labor value of all commodities will (weakly) fall \citet[Theorem~4.9]{roemer_1981}. Hence, a `larger' real wage bundle will still be compatible with a constant rate of exploitation. The `larger' real wage bundle can accommodate relatively higher magnitudes of commodities for which the labor values have fallen relatively more. If these commodities also had relatively high prices in the original situation compared to labor values after technical change, then the monetary cost of the real wage bundle will increase to such an extent that it will lead to a fall in the long run, equilibrium rate of profit.
	
	One important condition that ensures the fall in the equilibrium rate of profit is that the reduction in cost afforded by the new technique of production, evaluated with the original prices and the original real wage bundle, not be too large. In fact, if the cost reduction is bounded above by the change in the nominal labor cost associated with the new technique of production, then the equilibrium rate of profit will fall - squeezed by the rise in the nominal labor cost coming from the new real wage bundle \citet[Theorem~5]{dietzenbacher_1989}. Since new techniques of production are perturbations of current techniques \citep{dumenil_levy_1995}, the assumption of an upper bound on the cost reduction associated with a new technique of production seems reasonable.
	
	Given this bounded nature of cost reduction, the rate of profit falls because capitalists are unable to fully take account of the effects of technical change on the labor market. While capitalists might be able to control wage movements at the level of their firm, technical change has larger impacts on the labor market that is beyond the control of individual capitalists. It is this inability to fully control wage movements that, under certain plausible configurations of technological change, will lead to a fall in the long run, equilibrium rate of profit. Hence, individually rational capitalist actions can lead to an overall undermining of the interest of the whole capitalist class. 
	
	The rest of the paper is organized as follows. In section~\ref{sec:setup}, we describe the basic set up and define viable technical change. In section~\ref{sec:results}, we derive sufficient conditions for the rate of profit to fall after viable technical change if the rate of exploitation remains constant (these results are presented as Theorem~\ref{thm:frp}, and ~\ref{thm:existence}); subsequently, we show that if we impose a minor restriction on the permissible set of real wage bundles before technical change, then starting from any configuration of technology and real wage, there will always exist viable, CU-LS technical changes that will satisfy the sufficient conditions of Theorem~\ref{thm:existence} (this existence result is presented as Theorem~\ref{thm:existence-1}). In section~\ref{sec:example}, we present an example of a $3$ sector model to illustrate our argument; finally, we conclude the paper in section~\ref{sec:conclusion}.
	
	\section{The Set-Up}\label{sec:setup}
	
	\subsection{Initial Configuration}
	Consider an economy with $n$ sectors of production, where the technology is given by the non-negative $n \times n$ input-output matrix, $A \geqq 0$, and the $1 \times n$ vector of direct labor inputs, $L \gg 0$, and the real wage bundle given by the $n \times 1$ vector $b \geqq 0$.\footnote{For vectors and matrices, we will use the following notation: $x \geqq 0$, if for $i=1,2, \ldots, n$, $x_i \geq 0$ and $x \neq 0$; $x \gg 0$, if for $i=1,2, \ldots, n$, $x_i>0$.} Each sector produces one commodity with one technique of production and there is no fixed capital. 
	
	The cost of producing one unit of the commodity in sector $i$ is given by $p A_{*i}+wL_i$, where $A_{*i}$ denotes the $i$-th column of $A$, and $w=pb$ is the nominal wage rate. Using the normalization that the nominal wage rate is unity, the $1 \times n$ vector of long run equilibrium prices (prices of production), $p$, and the long run equilibrium (uniform) rate of profit, $\pi$, are given by
	\begin{equation}\label{eq:pop-1}
	p = \left( 1+\pi \right) p M, \textrm{ and } pb=1,
	\end{equation}
	where $M = A+bL$, is the augmented input matrix. 
	
	We assume that the input-output matrix, $A$, is productive and indecomposable. Then, an application of the Perron-Froebenius theorem shows that $p \gg 0$ and $\pi>0$ \citep[pp. 36]{dietzenbacher_1989}. For this configuration of technology, the $1 \times n$ vector of labor values, $\Lambda$, is given by
	\begin{equation}\label{value-def}
	\Lambda = L \left( I-A\right)^{-1}. 
	\end{equation}
	Standard results in linear algebra show that, since $A$ is productive, $(I-A)^{-1} \gg 0$ \citep[Appendix]{pasinetti_1977}. Hence, we have  $\Lambda \gg 0$. Once labor values and the real wage bundle is known, we can define the rate of exploitation as
	\begin{equation}
	e = \frac{1-\Lambda b}{\Lambda b}.
	\end{equation}
	
	\begin{assumption}\label{def-B}
		The real wage bundle, $b$, is an element of the set, $ \mathbb{B} = \mathbb{B}_1 \cap \mathbb{B}_2 $, where $\mathbb{B}_1 = \left\lbrace b \in \mathbb{R}^n_{+} \textrm{ s.t. } 0 < \Lambda b \leq 1\right\rbrace$, and $\mathbb{B}_2 =  \left\lbrace b \in \mathbb{R}^n_{+} \textrm{ s.t. } 1/(\Lambda b) = 1+e < \max_k  (p_k/\lambda_k)\right\rbrace$, where $p_k$ and $\lambda_k$ denote the $k$-th element of $p$ and $\Lambda$, respectively.
	\end{assumption}
	According to assumption~\ref{def-B}, a permissible real wage bundle before technical change must satisfy two restrictions. First, it must belong to $ \mathbb{B}_1$. This restriction is meant to rule out negative rates of exploitation, because the latter are conceptually meaningless. Note that there does not exist a one-one relationship between the rate of exploitation and the real wage bundle. In fact, given a positive rate of exploitation, $e>0$, any real wage bundle, $b$, that satisfies $\Lambda b = 1/(1+e)$ will be associated with $e$. The second condition in the definition of $\mathbb{B}$ imposes some restriction on the permissible real wage bundles and requires comment.
	
	The second restriction is that the real wage bundle, $b$, must belong to $\mathbb{B}_2$, i.e. $1+e = 1/(\Lambda b) < \max_k  (p_k/\lambda_k)$. This is a technical condition required for proving a result further down in the paper (Theorem~\ref{thm:existence-1}). It states that, given any positive rate of exploitation, $e>0$ (which is ensured by the first restriction), the real wage bundle must be such that the maximum price-value ratio among the $n$ commodities is strictly greater than the reciprocal of the value of the real wage bundle (which is equal to $1$ plus the rate of exploitation). This condition is less restrictive than might appear at first sight. \citet[Corollary~8.6]{roemer_1981} shows that, as long as the organic composition of capital is not identically equal in all sectors, $\max_k (p_k/\lambda_k) \geq 1+e \geq \min_k  (p_k/\lambda_k)$. Hence, the condition imposed in assumption~\ref{def-B} is just ruling out the possibility of an equality for the left hand weak inequality. We will comment on this condition below when we use it.
	
	\subsection{Viable, CU-LS Technical Change}
	Suppose there is a cost-reducing (viable) technical change in sector $i$, i.e. the cost of producing one unit of output with the new technique of production is lower than with the older technique of production when both are evaluated at current prices and wage rate. Hence,
	\begin{equation}\label{viability}
	p A_{*i} + L_i > p \bar{A}_{*i} + \bar{L}_i,
	\end{equation}
	where $A_{*i}$ and $\bar{A}_{*i}$ denote the $i$-th columns of the matrices $A$ and $\bar{A}$, respectively, $L_i$ denotes the $i$-th element of $L$, and we have used the normalization, once again, that the nominal wage rate is $1$. In addition, suppose technical change is capital-using and labor-saving (CU-LS). This means the the amount of material inputs used to produce one unit of the commodity in sector $i$ rises, while the amount of direct labor input falls. Hence, for $j = 1, 2, \ldots, n$,
	\begin{equation}\label{culs}
	a_{ji} < \bar{a}_{ji}, \textrm{ and } L_i > \bar{L}_i,
	\end{equation}
	where $a_{ij}$ and $\bar{a}_{ij}$ denote the $(i,j)$-th elements of $A$ and $\bar{A}$, respectively, and $L_i$ denotes the $i$-th element of $L$.
	
	Since the new technique of production reduces unit cost of production, evaluated at current prices, capitalist firms in sector $i$ will adopt the new technique of production \citep{okishio_1961}. All other sectors continue using the old technology - because technical change occurs only in sector $i$. Hence, the new technology in the economy is captured by the $n \times n$ input-output matrix, $\bar{A}$, and the $1 \times n$ vector of direct labor inputs, $\bar{L}$, where the columns of $A$ and $\bar{A}$ are identical other than for column $i$, and the elements of $L$ and $\bar{L}$ are identical, other than the $i$-th element. With the new technology, the $1 \times n$ vector of labor values, $\bar{\Lambda}$, is given by
	\begin{equation}
	\bar{\Lambda} = \bar{L} \left( I - \bar{A}\right)^{-1} \gg 0, 
	\end{equation}
	where strict inequality follows because $\bar{A}$ is productive.\footnote{We will only need viable technical change for the results in Theorem~\ref{thm:frp} and ~\ref{thm:existence} below; we will require CU-LS technical change for Theorem~\ref{thm:existence-1}.} 
	
	\subsection{Class Struggle and a New Real Wage Bundle}
	In this paper we will study the consequences of a viable, CU-LS technical change on the equilibrium rate of profit when the real wage bundle is allowed to vary. Suppose class struggle, during and after the technical change, leads to the emergence of a new real wage bundle, $\bar{b} \geqq 0$, with the following properties.
	
	\begin{property}\label{ass:more-exp}
	The new real wage bundle is such that $\bar{b} \in \mathbb{B}_1$ and
		\begin{equation}\label{ass:bexpense}
		p \bar{b}>pb=1.
		\end{equation}
	\end{property}
	This property specifies that the new real wage bundle is more expensive at original prices. This just means that workers are able to bargain for and secure a higher nominal wage rate as the process of technical change works itself out over the long run. 
	
	\begin{property}\label{ass:constexp}
		The new real wage bundle, $\bar{b}$, keeps the labor value of the real wage bundle unchanged, i.e.
		\begin{equation}
		\Lambda b = \bar{\Lambda} \bar{b}.
		\end{equation}
	\end{property}
	The rate of exploitation, before technical change, is given by $e=(1-\Lambda b)/\Lambda b$. After technical change and with the new real wage bundle, it is given by $\bar{e}=(1-\bar{\Lambda} \bar{b})/\bar{\Lambda} \bar{b}$. Hence, this property ensures that workers are able to secure a new real wage bundle that keeps the rate of exploitation unchanged even after technical change. If the rate of exploitation captures the balance of class forces, after taking account of technical change and its impact on the labor market, then this assumption states that there is no change in the balance of class forces over the course of technical change.
	
	\begin{property}\label{ass:costred}
		The decline in the unit cost of production in sector $i$ (the sector that witnessed technical change) is bounded above by the change in the nominal labor cost corresponding to the new technique of production,
		\begin{equation}\label{cond:costred}
		0 <p A_{*i} + L_i - p \bar{A}_{*i} - \bar{L}_i < \bar{L}_i \left( p \bar{b} - 1\right) . 
		\end{equation}
	\end{property}
	Since technical change in sector $i$ is viable, as captured by (\ref{viability}), it reduces the unit cost of production at current prices and wage rates. This gives us the left hand side of the inequality in (\ref{cond:costred}). The right hand side of of (\ref{cond:costred}), in addition, puts an upper bound on the decline in the unit cost of production. Note that the unit cost of production in sector $i$ before technical change is given by $p A_{*i} + L_i$; and, after technical change in that sector, it is given by $p \bar{A}_{*i} + \bar{L}_i$. Hence, $p A_{*i} + L_i - p \bar{A}_{*i} - \bar{L}_i$ is the decline in the unit cost of production in sector $i$. Since $pb=1$ was the nominal wage rate in the initial situation and $p \bar{b}$ is the nominal wage rate with the new real wage bundle, where both are evaluated at the original prices, $\bar{L}_i(p \bar{b} - 1)$ is the change in the nominal labor cost in sector $i$ corresponding to the direct labor input requirement associated with the new technique, $\bar{L}_i$. The right hand side of the inequality in (\ref{cond:costred}) states that the decline in unit cost of production is bounded above by $\bar{L}_i( p \bar{b} - 1)$.
	
	This condition is reasonable because technical change involves the emergence and adoption of new techniques of production that are local perturbations of the existing techniques of production \citep{dumenil_levy_1995}. Thus, while the amount of material and labor inputs required by the new technique is different from the old, the changes are not too large. The intuition of the change in cost corresponding to the new technique of production being not `too large' is captured by the above condition for the bound on the cost reduction associated with the new technique of production.
	
	\section{Main Results}\label{sec:results}
	The main results in this paper consists of three theorems. First, in Theorem~\ref{thm:frp} we show that if there exists some $\bar{b} \geqq 0$ that satisfies property~\ref{ass:more-exp}, ~\ref{ass:constexp} and ~\ref{ass:costred}, then viable technical change keeps the rate of exploitation constant even as the equilibrium rate of profit falls. This is a straightforward application of \citet[Theorem~5]{dietzenbacher_1989}.
	
	Second, in Theorem~\ref{thm:existence} we derive sufficient conditions for any new real wage bundle $\bar{b} \in \mathbb{B}_1$ to satisfy properties~\ref{ass:constexp} and ~\ref{ass:costred}. The implication is that if class struggle leads to the emergence of such a real wage bundle then viable technical change will be accompanied by a fall in the uniform rate of profit even as the rate of exploitation remains unchanged. Analytically, the main challenge is to show that such a real wage bundle exists and is economically meaningful. While it is easy to see that an increase in the real wage bundle will lead to a fall in the equilibrium rate of profit, it is not immediately obvious that such a real wage bundle can also keep the rate of exploitation unchanged. The second theorem below provides sufficient conditions for the existence of such real wage bundles.\footnote{Note that for the first two theorems, we do not need the assumption of CU-LS technical change. We only need technical change to be viable, i.e. cost-reducing at original prices.} 
	
	While theorem~\ref{thm:existence} takes us some distance, it still leaves the question of existence unaddressed. That is, we still need to ask if, starting from any configuration of technology and real wage (that satisfies assumption~\ref{def-B}), we can find a new viable technique of production that satisfies the sufficient condition of the second theorem. Theorem~\ref{thm:existence-1} in this paper answers this question in the affirmative for the class of CU-LS technical change. In this theorem, we show that, starting from \textit{any} configuration of technology and the real wage bundle (that satisfies assumption~\ref{def-B}), there always exists some viable, CU-LS technological change that satisfy the sufficient condition of the theorem~\ref{thm:existence}. The three theorems together show that for any configuration of technology and real wage bundle (that satisfies assumption~\ref{def-B}), it is always \textit{possible} for a capitalist economy to witness a fall in the equilibrium rate of profit alongside a constant rate of exploitation, after a viable, CU-LS technical change.

	\begin{theorem}\label{thm:frp}
		Let $p$ and $\pi$ denote the price of production vector and the uniform rate of profit with the technology given by $A, L$ and the real wage bundle given by $b \in \mathbb{B}$. Suppose there is a viable technological change, with the new technology given by $\bar{A}, \bar{L}$ and the real wage bundle given by $\bar{b} \in \mathbb{B}_1$. Let $\bar{p}$ and $\bar{\pi}$ denote the price of production vector and the uniform rate of profit with the new technology. If the new real wage bundle $\bar{b}$ satisfies property~\ref{ass:more-exp}, ~\ref{ass:constexp} and ~\ref{ass:costred}, then $\bar{\pi}<\pi$. 
	\end{theorem}
	\begin{proof}
		Since $\bar{b}$ satisfies property~\ref{ass:constexp}, the rate of exploitation remains unchanged. An application of \citet[Theorem~5]{dietzenbacher_1989} shows that, since (\ref{ass:bexpense}) and (\ref{cond:costred}) hold, the uniform rate of profit declines.
	\end{proof}
	
	The implication of this result is interesting. It shows that if a more expensive real wage bundle satisfying property~\ref{ass:constexp} and ~\ref{ass:costred} exists, then viable technical change can, at the same time, keep the rate of exploitation constant and also lead to a fall in the uniform rate of profit. Hence, this shows that Marx's claim in Volume III of \textit{Capital} can be sustained under certain conditions. Of course, to complete the argument, we must demonstrate that such a real wage bundle $\bar{b}$ actually exists. In the next theorem we provide sufficient conditions for the existence of such a real wage bundle.

\begin{figure}
		\centering
		\begin{subfigure}[b]{0.4\textwidth}
			\centering
			\includegraphics[width=\textwidth]{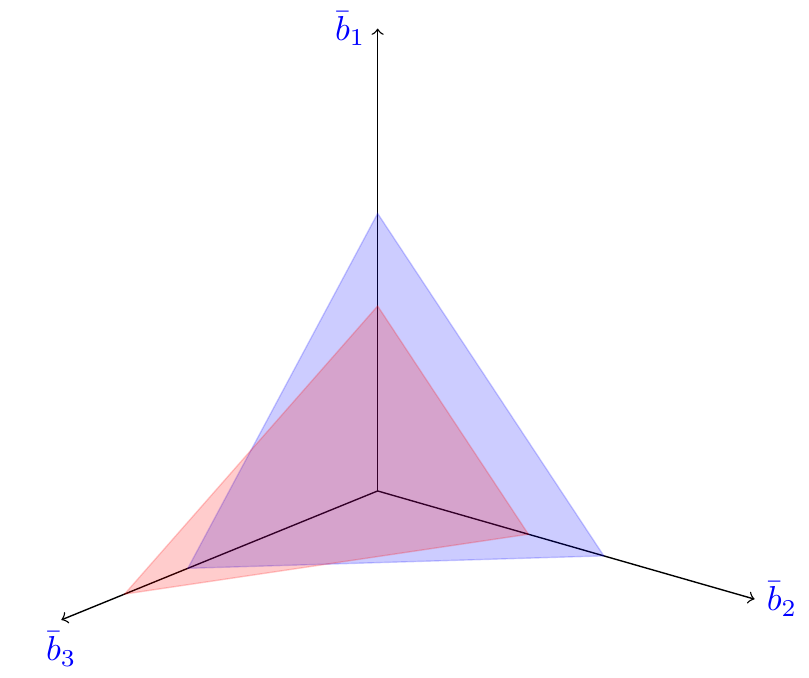}
			\caption{The hyperplanes $P$ (blue, extending outward on the $\bar{b}_1$ axis) and $V$ (red, extending outward on the $\bar{b}_3$ axis).}
			\label{fig:simplex}
		\end{subfigure}
		\hfill
		\begin{subfigure}[b]{0.4\textwidth}
			\centering
			\includegraphics[width=\textwidth]{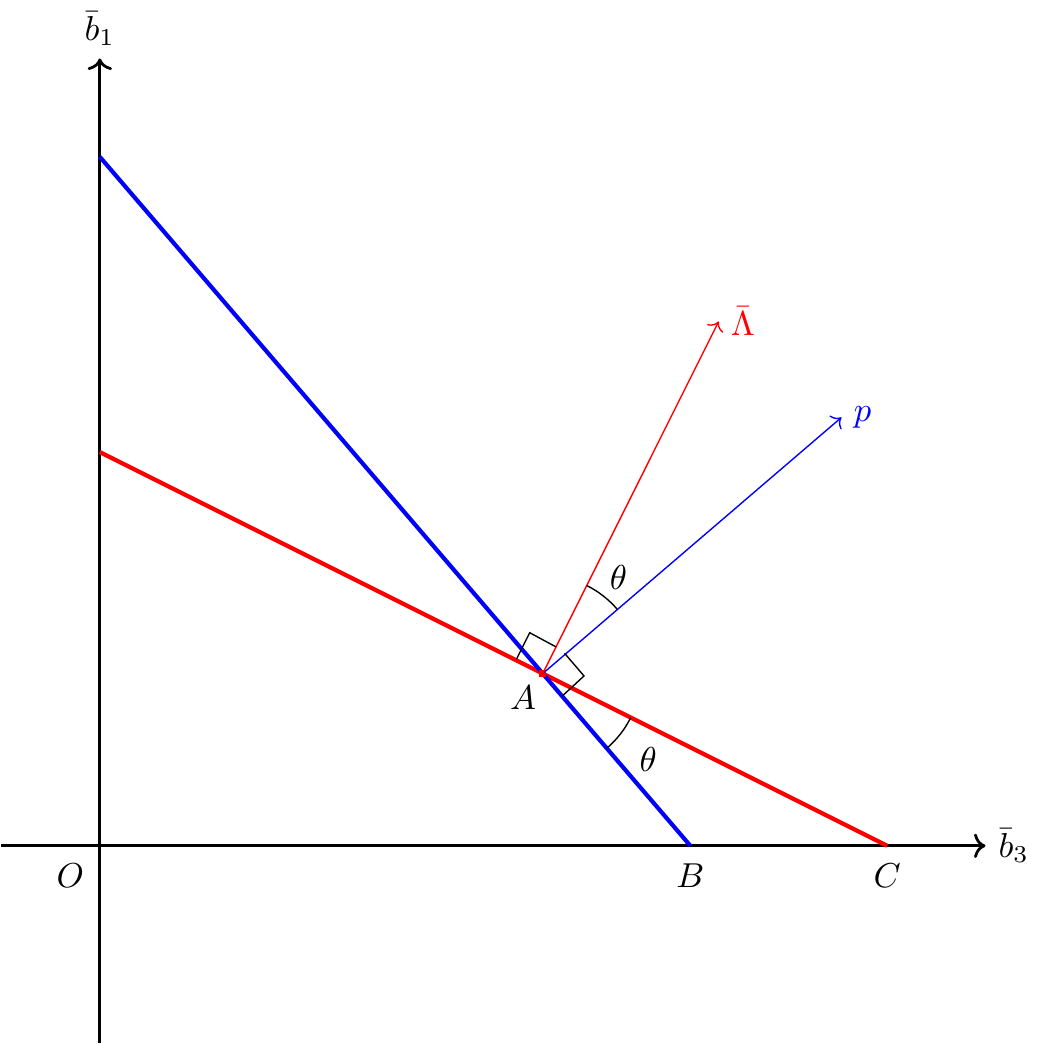}
			\caption{Projection of the hyperplanes $P$ (blue) and $V$ (red) onto the plane defined by $\bar{b}_2=0$.}
			\label{fig:2d}
		\end{subfigure}
		\caption{Intersecting hyperplanes given by equations (\ref{defP}) and (\ref{defV}) for the case $n=3$.}
		\label{fig:hyper}
	\end{figure}

	\begin{theorem}\label{thm:existence}
		Let $e$ denote the rate of exploitation before technical change, i.e.
		\begin{equation}\label{def:exp}
		e = \frac{1-\Lambda b}{\Lambda b}>0,
		\end{equation}
		let $p$ denote the initial price of production vector, and let $g$ denote the decline in the cost of production in sector $i$ (the sector which witnessed viable technical change) as a fraction of the labor cost corresponding to the new technique of production in that sector evaluated at the old wage rate,
		\begin{equation}\label{defg}
		g = \frac{\left( p A_{*i} + L_i\right) - \left( p \bar{A}_{*i} + \bar{L}_i\right)}{\bar{L}_i}>0.
		\end{equation}
		Let $\bar{\Lambda}=[\bar{\lambda}_i]$ denote the vector of labor values after the viable technical change described in Theorem~\ref{thm:frp}. If, for some $j=1,2, \ldots, n$, 
		\begin{equation}\label{thm:cond}
		\left( p_j/\bar{\lambda}_j\right)  > \left( 1+e\right)\left( 1+g\right),  
		\end{equation}
		then there exists some $\bar{b} \in \mathbb{B}_1$ that satisfies property~\ref{ass:more-exp}, ~\ref{ass:constexp} and ~\ref{ass:costred}.
		
	\end{theorem}
	\begin{proof}
		Consider the $n$ dimensional space whose coordinate system is ($\bar{b}_1, \ldots, \bar{b}_n$). Any point in this space is a candidate real wage bundle. We will only consider the nonnegative orthant of this space because negative elements in the real wage bundle are not meaningful.
		
		Consider the hyperplane in this space given by the set of points $P$ defined by
		\begin{equation}\label{defP}
		P = \left\lbrace \bar{b} \geqq 0 \quad | \quad p \cdot \bar{b} - \alpha = 0 \right\rbrace, 
		\end{equation}
		where $i$ denotes the sector in which technical change occurred, and 
		\begin{equation}\label{defalpha}
		\alpha = ( p A_{*i} + L_i - p \bar{A}_{*i} )/\bar{L}_i>1,
		\end{equation}
		where the strict inequality in (\ref{defalpha}) comes from the fact that $\alpha = 1+g$ and $g>0$. Note that, for $j=1, 2, \ldots, n$, the hyperplane $P$ intersects the coordinate axes at points of the form $x_j e_j$, where $e_j$ is a $n$-vector with $1$ as the $j$-th element and $0$ as every other element, and $ x_j = \alpha/p_j > 0 $, where the strict inequality follows from (\ref{defalpha}) and $p_j>0$. Thus, the hyperplane $P$ intersects the coordinate axes at strictly positive points. The important point to note is that all points `above' hyperplane, $P$, satisfies property~~\ref{ass:more-exp} and ~\ref{ass:costred}, i.e. such real wage bundles are more expensive at original prices and the decline in the unit cost of production in sector $i$ is bounded above by the change in labor cost corresponding to the new technique of production.
		
		Now consider the hyperplane, in the same $n$ dimensional space, given by the set of points $V$ defined by  
		\begin{equation}\label{defV}
		V = \left\lbrace \bar{b} \geqq 0 \quad | \quad \bar{\Lambda} \cdot \bar{b} - \beta = 0 \right\rbrace, 
		\end{equation}
		where 
		\begin{equation}\label{defbeta}
		\beta = \Lambda b>0,
		\end{equation}
		where the strict inequality in (\ref{defbeta}) follows because $\Lambda>0$ and $b\geq 0$. Note that, for $j=1, 2, \ldots, n$, this hyperplane intersects the coordinate axes at points of the form $y_j e_j$ where, $y_j = \beta/\bar{\lambda}_j > 0$, where the strict inequality follows from (\ref{defbeta}) and $\bar{\lambda}_j>0$. Thus, the hyperplane $V$ also intersects the coordinate axes at strictly positive points. For the hyperplane, $V$, the important point to note is that all points on this hyperplane satisfy property~\ref{ass:constexp}, i.e. such real wage bundles ensure that the rate of exploitation remains unchanged before and after technical change.
		
		Using the assumption of the theorem, condition (\ref{thm:cond}), we have, for at least one $j=1, 2, \ldots, n$,  $ ( p_j/\bar{\lambda}_j)  > ( 1+e)( 1+g)$. Using (\ref{defg}), we see that $(1+g)=(p A_{*i}+L_i-p \bar{A}_{*i})/\bar{L}_i$. Using (\ref{def:exp}), we also know that $(1+e)=(1/\Lambda b)$. Using these, we have,
		\begin{align*}
		\frac{p_j}{\bar{\lambda}_j} & > ( 1+e)( 1+g) = \frac{1}{\Lambda b}\left[ \frac{p A_{*i}+L_i-p \bar{A}_{*i}}{\bar{L}_i}\right] = \frac{1}{\Lambda b}\left[ \frac{p A_{*i}+L_i-p \bar{A}_{*i}}{\bar{L}_i}\right] = \frac{ \alpha}{\beta},
		\end{align*}
		so that $\beta/\bar{\lambda}_j>\alpha/p_j$. This means that some portion of the hyperplane, $V$, lies above the hyperplane, $P$, in the \textit{positive orthant}, as shown in Figure~\ref{fig:simplex} and ~\ref{fig:2d} for a $3$-dimensional setting. This provides us with an infinite number of points $\bar{b} \in \mathbb{B}_1$ for which property~\ref{ass:more-exp}, ~\ref{ass:constexp} and ~\ref{ass:costred} will be satisfied. To show this more formally, we need to demonstrate that points on the hyperplane $V$ that lie in the positive orthant \textit{and} to the right of the intersection with hyperplane $P$ are `above' the hyperplane $P$. 
		
		Let  $j=k$ for which (\ref{thm:cond}) is satisfied, i.e. $ (p_k/\bar{\lambda}_k)  > ( 1+e)( 1+g)$. Consider the hyperplane given by the set of points $B_{k-1}$ defined by $B_{k-1} = \left\lbrace \bar{b} \geqq 0 | \bar{b}_{k-1}=0 \right\rbrace$. Let $x$ denote the point of intersection of $P$, $V$ and $B_{k-1}$. Let $y$ denote the point where the hyperplane $V$ intersects the $b_k$ coordinate axis, and let $z$ denote the point where the hyperplane $P$ intersects the $b_k$ coordinate axis. Let $u$ denote the unit vector given by $u = (y-x)/\|y-x\|$, and let $v$ denote the unit vector given by $v = (z-x)/\|z-x\|$, where $\|w\|$ denotes the Euclidean norm of the vector $w$. The vector ($y-x$) lies on the hyperplane $V$, and the vector ($z-x$) lies on the hyperplane $P$.\footnote{In Figure~\ref{fig:2d}, $x=OA, y=OC, z=OB$. Hence, $y-x=CA$ and $z-x=BA$. } Hence, the angle between $u$ and $v$ is the angle between $p$ and $\bar{\Lambda}$, because the vector $p$ is perpendicular to the hyperplane $P$ and the vector $\bar{\Lambda}$ is perpendicular to hyperplane $V$ (see Figure~\ref{fig:2d}).\footnote{Recall that the $P$ is given by $p \cdot \bar{b} = \alpha$, and $V$ is given by $\bar{\Lambda} \cdot \bar{b} = \beta$. This shows that the vector $p$ is perpendicular to $P$ and the vector $\bar{\Lambda}$ is perpendicular to $V$.} Hence, if $\theta$ denotes the angle between $u$ and $v$, then $\cos \theta = (p \cdot \bar{\Lambda})/\|p\|\|\bar{\Lambda}\|>0$, where the strict inequality comes from the fact that both $p$ and $\bar{\Lambda}$ are strictly positive vectors.
		
		Let $0<t<1$, and consider a point $x_0 = x + tu$, and note that $x_0$ lies on $V$. We will show that $x_0$ is `above' $P$ by showing that it lies on the other side of $P$, compared to the origin. When we plug in the zero vector (origin) in the equation for the hyperplane $P$, we get $-\alpha<0$. When we plug in $x_0$ into the equation for the same hyperplane $P$, we get $p \cdot x+tp \cdot u$. Since $x$ lies on $P$, we have $p \cdot x = \alpha>0$. Additionally, $tp \cdot u = t \|p\| \cos \theta >0$, because $\cos \theta>0$ (as we saw above). Hence, $p \cdot x+tp \cdot u>0$. This shows that the point $x_0$ is `above' the hyperplane $P$, i.e. the origin and $x_0$ are on \textit{different} sides of the hyperplane $P$.		 
	\end{proof}
	
	\textit{Discussion.} The key condition in Theorem~\ref{thm:existence} is captured in (\ref{thm:cond}). This instructs us to look at the ratio of the price of production before technical change and the labor value after technical change, sector by sector. If for any sector, this ratio is bounded below by the product $(1+e)(1+g)$, then the condition will be satisfied. When technical change is CU-LS, this condition is not restrictive, because $p\gg \Lambda \geqq \bar{\Lambda}$, i.e. the vector of prices of production before technical change is strictly larger than the vector of values after technical change.

	To see the first inequality, $p \gg \Lambda$, note that, from (\ref{eq:pop-1}), we have, $p = (1+\pi)L[I - (1+\pi)A]^{-1} = (1+\pi)L \sum_{j=1}^{\infty}(1+\pi)^jA^j \gg (1+\pi)L \sum_{j=1}^{\infty}A^j = (1+\pi)L[I - A]^{-1}= (1+\pi)\Lambda \gg \Lambda$, where we have used the facts that $0<\pi<R$ (where $R$ is the maximal rate of profit, i.e. the rate of profit when the real wage bundle is the zero vector) and $A$ is productive. As long as the real wage bundle, $b \in \mathbb{B}$, has at least one strictly positive element, we have $0<\pi<R$, where $1+R$ is the reciprocal of the maximal eigenvalue of $A$, and this ensures the validity of the infinite series matrix expansion of $ [I - (1+\pi)A]^{-1} $; and, as long as $A$ is productive, we have a valid infinite series matrix expansion for $[I - A]^{-1}$.\footnote{Note, we have used the notation $A^0=I$, the $0$-th power of the matrix $A$ is the identity matrix. For a discussion of this infinite series matrix expansion, see \citet[Appendix, pp. 266]{pasinetti_1977}.} The second inequality, $\Lambda \geqq \bar{\Lambda}$, follows from \citet[Theorem~4.9]{roemer_1981} because the technical change under consideration is CU-LS.
	
	Given these inequalities, i.e. $p \gg \Lambda \geqq \bar{\Lambda}$, the condition in (\ref{thm:cond}) is merely stating that the ratio of price of production (before technical change) and the labor value (after technical change) in at least one sector must not only by larger than unity, but be larger than the quantity appearing on the right hand side of (\ref{thm:cond}). The intuition behind this condition is that, since CU-LS technical change reduces the labor value, we can increase the magnitude of commodities in the real wage bundle with relatively low labor value and yet keep the value of the wage bundle unchanged. But, if these commodities had relatively high prices in the original situation compared to their labor values after technical change, then the monetary cost of the real wage bundle increases to such an extent that it leads to a fall in the equilibrium rate of profit.
	
	While this intuitive argument might be persuasive, it still leaves open the question of a rigorous demonstration of existence. Starting from \textit{any} configuration of technology and real wage bundle (that satisfies assumption~\ref{def-B}), can we always find some viable, CU-LS technological change that satisfies the condition in (\ref{thm:cond})? The next result shows that this question can be answered in the affirmative.
	
	\begin{theorem}\label{thm:existence-1}
		Let $A$ and $L$ denote any input-output matrix and direct labor input vector, respectively, and let the real wage bundle be given by, $b \in \mathbb{B}$, as defined in assumption~\ref{def-B}. Then, a new input-output matrix, $\bar{A}$, and a direct labor input vector, $\bar{L}$, exists such that the new technology is viable, CU-LS and the condition in (\ref{thm:cond}) is satisfied.
	\end{theorem}
	\begin{proof}
		Given $A$, and $L$, we can calculate the vector of labor values, $\Lambda$, using (\ref{value-def}). Choose a real wage bundle, $b \in \mathbb{B}$, defined in assumption~\ref{def-B}. The definition of $\mathbb{B}$ ensures that $\max_k (p_k/\lambda_k)> (1+e) = (1/\Lambda b)$. Let  $j$ be the index for which the value-price ratio attains the maximum magnitude. Let $\phi=(\Lambda b p_j)/\lambda_j$, and note that $\phi>1$. 
		
		
		Let $i$ denote the industry in which technological change occurs. Choose $\varepsilon$ such that such that $0<\varepsilon<L_i/\sum_k p_k$, which ensures that $L_i - \varepsilon \sum_k p_k>0$. Now construct the new technique of production for industry $i$ as follows: choose $\bar{L}_i$ such that $(L_i- \varepsilon \sum_k p_k)/\phi < \bar{L}_i<L_i-\varepsilon \sum_k p_k$ (which is always possible because $\phi > 1$\footnote{It is precisely here that the second property in assumption~\ref{def-B} is used. We need $\phi>1$ as a strict inequality because if $\phi=1$, then $(L_i- \varepsilon \sum_k p_k)/\phi = L_i-\varepsilon \sum_k p_k$ and we would not be able to choose a $\bar{L}_i$. It is to rule out this eventuality that we must ensure the strict inequality in the second condition defining the set $\mathbb{B}$ in assumption~\ref{def-B}.}), and construct the vector $\bar{A}_{*i}$ by adding $\varepsilon$ to \textit{every} element of the vector $A_{*i}$. For all other industries, the technique of production remains unchanged. We will now show that this new technology given by, $\bar{A}, \bar{L}$, is CU-LS, viable, and satisfies (\ref{thm:cond}).
		
		Since every element of the vector $\bar{A}_{*i}$ is strictly greater than the corresponding element of the vector $A_{*i}$ and $\bar{L}_i<L_i-\varepsilon \sum_k p_k<L_i$, the new technology is CU-LS. To see viability, let us use the right hand inequality defining $\bar{L}_i$: $ \bar{L}_i<L_i-\varepsilon \sum_k p_k $. Since $p \bar{A}_{*i} - p A_{*i} = \varepsilon \sum_k p_k$, we have $\bar{L}_i< L_i-\varepsilon \sum_k p_k=L_i-(p \bar{A}_{*i} - p A_{*i})$, and viability is established.
		
		Let us now use the left hand inequality defining $\bar{L}_i$: $(L_i- \varepsilon \sum_k p_k)/\phi < \bar{L}_i$. This guarantees that $L_i - \phi \bar{L}_i < \varepsilon \sum_k p_k = p (\bar{A}_{*i}-A_{*i})$. Using the definition of $\phi$, this gives us $(p A_{*i}+L_i - p \bar{A}_{*i})/\bar{L}_i < \Lambda b p_j/\lambda_j$. Hence, we have, 
		\[
		\frac{1}{\Lambda b}\left[ \frac{p A_{*i}+L_i - p \bar{A}_{*i}}{\bar{L}_i}\right] < \frac{p_j}{\lambda_j}. 
		\]
		
		Since the new technology is CU-LS, \citet[Theorem~4.9]{roemer_1981} shows that $0<\bar{\lambda}_j \leq \lambda_j$. Hence, $p_j/\lambda_j \leq p_j/\bar{\lambda}_j$. Thus,
		\[
		\frac{1}{\Lambda b}\left[ \frac{p A_{*i}+L_i - p \bar{A}_{*i}}{\bar{L}_i}\right] < \frac{p_j}{\lambda_j} \leq \frac{p_j}{\bar{\lambda}_j}, 
		\]
		so that condition (\ref{thm:cond}) is satisfied.
	\end{proof}
	
	\textit{Discussion.} The result in theorem~\ref{thm:existence-1} shows that, starting from any configuration of technology and real wage bundle (that satisfies assumption~\ref{def-B}), we can always find a new technology given by $\bar{A}$ and $\bar{L}$ such that the technology is viable, CU-LS and condition (\ref{thm:cond}) is satisfied. Taken together, theorems~\ref{thm:frp}, ~\ref{thm:existence} and ~\ref{thm:existence-1} show that, starting from any configuration of technology and distribution (real wage bundle), a capitalist economy can always witness a viable, CU-LS technical change that keeps the rate of exploitation constant and leads to a fall in the uniform rate of profit. This demonstrates that Marx's claim in Volume III of \textit{Capital} that the rate of profit can fall due to technical change if the rate of exploitation remains unchanged can be sustained in certain plausible configurations of technology and distribution. 
	
	The requirement that the rate of exploitation remain unchanged over the period of technical change, i.e. as the economy moves from the old to the new long run equilibrium, is restrictive. Moreover, it contradicts another important claim that Marx developed in Volume I of \textit{Capital}, that the rate of exploitation rises with the development of capitalism, in the form of the production of absolute or relative surplus value. The fact of a rising rate of exploitation, which is less restrictive than the assumption of a constant rate of exploitation, can be accommodated in our framework.
	
	\begin{corollary}\label{thm:exp-rise}
		Starting from any configuration of technology and distribution (real wage bundle), a capitalist economy can always witness a viable, CU-LS technical change that allows the rate of exploitation to increase and leads to a fall in the uniform rate of profit.
	\end{corollary}
	\begin{proof}
		The proof follows by noting that as long as the sufficient condition in theorem~\ref{thm:existence} is satisfied, there will exist real wage bundles which are below the hyperplane $V$ and above the hyperplane $P$ (e.g. points in the interior of triangle $ABC$ in Figure~\ref{fig:2d}). For such real wage bundles, the rate of exploitation will rise (because they are below $V$) and the rate of profit will fall (because they are above $P$).
	\end{proof}
	
	Corollary~\ref{thm:exp-rise} shows that we \textit{can} have a rise in the rate of exploitation, in the form of the production of relative surplus value, and yet viable, CU-LS technical change can lead to a fall in the rate of profit. Hence, the interaction of class struggle and technical change can allow for the rate of exploitation to rise, and yet the equilibrium rate of profit might decline under the conditions laid out in theorem~\ref{thm:existence}. 
	
	In the next section, we provide an example of a $3$-sector economy where we can find an infinite number of real wage bundles that, after a viable, CU-LS technical change, can keep the rate of exploitation constant and lead to a fall in the equilibrium rate of profit. But a caveat is necessary at this point. Our argument is not that class struggle will always discover a real wage bundle that satisfies property~\ref{ass:constexp} and ~\ref{ass:costred}. Rather, we have demonstrated that such a real wage bundle does exist and that it is not possible to rule it out without additional restrictions on technical change or class struggle. Hence, it is \textit{possible} that such a real wage bundle will be discovered by class struggle. In that case, the equilibrium rate of profit will fall even when capitalists have adopted cost-reducing techniques of production.

	\section{An Example}\label{sec:example}
	Consider a slight variation in the example of a $3$-sector economy discussed in \citet[pp. 39]{dietzenbacher_1989}.\footnote{R code to implement this example is given in the Appendix.} 	
	
	\subsection{Initial Situation}
	Let initial technology be given by
	\[
	A = \begin{bmatrix}
	0.35 & 0.05 & 0.25 \\
	0.15 & 0.45 &  0.05 \\
	0.15 & 0.15 & 0.35
	\end{bmatrix}
	\]
	and
	\[
	L = \begin{bmatrix}
	0.2 & 0.15 & 0.25
	\end{bmatrix}.
	\]
	Let the initial real wage bundle be given by 
	\[
	b = \begin{bmatrix}
	1/3 \\
	1/3 \\
	1/3
	\end{bmatrix}. 
	\]
	
	For this configuration of technology and real wage bundle, we can calculate the uniform rate of profit, $\pi=0.17647$, and the price of production vector as
	\[
	p = \begin{bmatrix}
	1 & 0.9090909 & 1.090909
	\end{bmatrix}.
	\]
	The vector of values is given by
	\[
	\Lambda = \begin{bmatrix}
	0.5714286  & 0.5 & 0.6428571
	\end{bmatrix}.
	\]
	Hence, $pb=1$ and $\Lambda b = 0.5714286$. 
	
	Let us check that the chosen real wage bundle satisfies the two properties specified in assumption~\ref{def-B}. Since $\Lambda b = 0.5714286<1$, the first property is satisfied. Moreover, $1/\Lambda b = 1.75$, and $\max_k (p_k/v_k)=1.8182$. Hence, the second condition is satisfied.
	
	\subsection{CU-LS, Viable Technical Change}
	A CU-LS, viable technical change takes place in sector 3. The new technology is given by
	\[
	\bar{A} = \begin{bmatrix}
	0.35 & 0.05 & 0.27 \\
	0.15 & 0.45 &  0.07 \\
	0.15 & 0.15 & 0.37
	\end{bmatrix}
	\]
	and
	\[
	\bar{L} = \begin{bmatrix}
	0.2 & 0.15 & 0.18
	\end{bmatrix}.
	\]
	Note that the new technology is  
	\begin{itemize}
		\item CUS-LS: because the third column of $\bar{A}$ is, element by element, greater than the third column of $A$, and the third element of $\bar{L}$ is strictly less than the third element of $L$;
		\item viable: because the cost of production in sector 3 falls from $0.9272727$ to $0.9172727$ (using the price vector computed above and using the normalisation that the nominal wage rate is $1$).
	\end{itemize}
	Hence, a capitalist producer will adopt this new technology. With this new technology, the new vector of value is given by
	\[
	\bar{\Lambda} = \begin{bmatrix}
	0.5511364 & 0.4797078 & 0.5752165
	\end{bmatrix}.
	\]
	
	\subsection{Class Struggle and a New Real Wage Bundle}
	Suppose class struggle leads to the emergence of a new real wage bundle given by
	\[
	\bar{b} = \begin{bmatrix}
	\bar{b}_1 \\
	\bar{b}_2 \\
	\bar{b}_3
	\end{bmatrix}. 
	\]
	
	We need to ensure that the new real wage vector $\bar{b} \geq 0$ is more expensive than the original real wage bundle
	\begin{equation}
	p_1 \bar{b}_1 + p_2 \bar{b}_2 + p_3 \bar{b}_3 > 1=pb, 
	\end{equation}
	and that the decline in the unit cost of production in bounded above by the change in the nominal labor cost associated with the new technique of production,
	\begin{equation}
	pA_{*3} + L_i < p\bar{A}_{*3} + (p_1 \bar{b}_1 + p_2 \bar{b}_2 + p_3 \bar{b}_3) \bar{L}_3, 
	\end{equation}
	where $A_{*3}$ and $\bar{A}_{*3}$ denote the third column of $A$ and $\bar{A}$, respectively, and, finally, that the labor value of the real wage bundle remains unchanged,
	\begin{equation}
	\bar{\lambda}_1 \bar{b}_1 + \bar{\lambda}_2 \bar{b}_2 + \bar{\lambda}_3 \bar{b}_3 = \Lambda b, 
	\end{equation}
	which ensures that the rate of exploitation remains constant.
	
	Since the new technique of production reduces the unit cost of production in sector $ 3 $, we have $(pA_{*3} + L_i - p\bar{A}_{*3})/\bar{L}_3>1$. Hence, the above three conditions can be reduced to two conditions:
	\begin{align}
	p_1 \bar{b}_1 + p_2 \bar{b}_2 + p_3 \bar{b}_3 & > (pA_{*3} + L_i - p\bar{A}_{*3})/\bar{L}_3, \label{cond1}\\
	\bar{\lambda}_1 \bar{b}_1 + \bar{\lambda}_2 \bar{b}_2 + \bar{\lambda}_3 \bar{b}_3 = \Lambda b. \label{cond2} 
	\end{align}

	Since $(pA_{*3} + L_i - p\bar{A}_{*3})/\bar{L}_3=1.055556$, and using the vector of old price of production, the vector of old and new labor values, we have the following two equations:
	\begin{align}
	1*\bar{b}_1 + 0.9090*\bar{b}_2 + 1.0909*\bar{b}_3 & > 1.055556 ,\label{cond11}\\
	0.5511364 \bar{b}_1 + 0.4797078 \bar{b}_2 + 0.5752165 \bar{b}_3 & = 0.5714286 \label{cond21}.
	\end{align}

	The condition in (\ref{cond11}) is satisfied by all points in the positive orthant of the 3-dimensional space with coordinates $(\bar{b}_1,\bar{b}_2,\bar{b}_3)$ that lie above the hyperplane $\bar{b}_1 + 0.9090*\bar{b}_2 + 1.0909*\bar{b}_3 = 1.055556$. This hyperplane intersects the three axes at $(1.0555556,0,0)$, $(0,1.1611111,0)$, and $(0,0,0.9675926)$. The condition is (\ref{cond21}) is satisfied by all points in 3-dimensional space with coordinates $(\bar{b}_1,\bar{b}_2,\bar{b}_3)$ that lie on the hyperplane given by (\ref{cond21}). This hyperplane intersects the three axes at $(1.0368189,0,0)$, $(0,1.1912014,0)$, and $(0,0,0.9934149)$. Hence, the two hyperplanes will intersect in the positive orthant. Thus, there are an infinite number of points that lie on the hyperplane given by (\ref{cond21}) and that also satisfy (\ref{cond11}).
	
	To choose one \textit{particular} real wage bundle, $\bar{b} \geq 0$, that satisfies (\ref{cond11}) and (\ref{cond21}), let us draw from a uniform distribution with support on $(1.1611111, 1.1912014)$. The draw gives us the second element of the new real wage bundle: $\bar{b}_2=1.170977$. Let us also impose the condition $\bar{b}_1=\bar{b}_3$ to simplify the computation. Hence, using (\ref{cond21}), we get $\bar{b}_1=(0.5714286 - 1.170977*0.4797078)/(2*0.5752165)$. Hence, $\bar{b}_1=0.008613$. Hence, the new real wage bundle is given by  
	\[
	\bar{b} = \begin{bmatrix}
	0.008613 \\
	1.170977 \\
	0.008613
	\end{bmatrix}. 
	\]
	Note that the equilibrium rate of profit now becomes $ \bar{\pi}=0.1604551 $, and the new price of production vector is given by 
	\[
	\bar{p} = \begin{bmatrix}
	0.9288424 & 0.8398318 & 0.9956171
	\end{bmatrix}.
	\]
	Hence, $\max_k (\bar{p}_k/\bar{\lambda}_k)=1.750715 > 1.75 = 1/(\bar{\Lambda}\bar{b})$. Thus, the condition for assumption~\ref{def-B} is satisfied by the new real wage bundle.

	Note, finally, that using this real wage bundle, we get a constant value of the real wage bundles, $\Lambda b = \bar{\Lambda} \bar{b}=0.5714286$. This keeps the rate of exploitation constant. Moreover, the new uniform rate of profit is $\bar{\pi}=0.1604551<0.1764706=\pi$. Hence, the equilibrium rate of profit falls after a viable, CU-LS technical change.

	\section{Conclusion}\label{sec:conclusion}
	Technical change is a characteristic feature of capitalist economies. Since the profit rate is one of the clearest indicators of the health of a capitalist economy, seen from the perspective of capital, it is of great interest to investigate the effect of technical change on the rate of profit. In Volume III of \textit{Capital}, Marx had argued that technical change will impart a falling tendency to the rate of profit when the rate of exploitation remains constant. In this paper, we have demonstrated that this result \textit{can} be obtained in a multisector economy. To be more concrete, we have demonstrated that there exist plausible real wage bundles which keep the rate of exploitation constant and lead to a fall in the new equilibrium rate of profit even after viable, CU-LS technical change. 
	
	The picture of technical change that Marx gave us in the volumes of \textit{Capital} remains extremely relevant. Competitive pressures force capitalists to search for new cost-reducing techniques of production. The innovator capitalist who manages to adopt such a technique is able to make super-normal profits. That creates the incentive for capitalists to constantly look for and adopt, when found, cost-reducing techniques of production. The adoption of the new technique by the innovator disrupts the prevailing equilibrium. This is because the economy is interconnected in complex ways. The output of the innovator capitalist is used as inputs in other industries; the innovator's demand for the output of other industries also change because of the technical change. When all these changes have played themselves out, a new equilibrium profit rate and a new set of prices of production emerge.
	
	\citet{okishio_1961} had shown that the new equilibrium rate of profit would be higher than the one that prevailed before technical change if the real wage rate remains unchanged. In this paper, we have shown that if the rate of exploitation remains unchanged, which will imply that the real wage rate has to increase, the rate of profit can fall after cost-reducing technical change of the type analysed by \citet{okishio_1961}, as long as the reduction in cost is bounded above by the change in the nominal labor cost associated with the new technique of production. The constancy of the rate of exploitation is one way to capture the balance of class forces. Hence, the result in this paper shows that if the balance of class forces manages to keep the division between paid and unpaid labour time unchanged, cost-reducing technical change can lead to a fall in the rate of profit - if the cost reduction from technical change is not too large. In such cases, individually rational decisions by capitalist producers might harm the collective interest of the capitalist class. This is just one pathology of a competitive, capitalist economy.

	\bibliography{/home/basu15/Dropbox/Marx-Economics/Manuscript/marxrefs}
	
	\begin{appendices}
		
		\section{R Code for the Example}
		
		\begin{verbatim}
		# Input-output data
		A <- matrix(c(0.35,0.05,0.25,
		0.15, 0.45, 0.05,
		0.15, 0.15, 0.35), 
		byrow = TRUE, ncol = 3, nrow = 3)
		
		# See A matrix
		A
		
		# Direct labor input
		l <- matrix(c(0.2,0.15,0.25),ncol=3)
		
		# See l vector
		l
		
		# Real wage bundle
		b <- matrix(c(1/3,1/3,1/3),ncol = 1)
		
		# See real wage bundle
		b
		
		# Augmented input matrix
		M <- A + b%*%l
		
		# See M matrix
		M
		
		# Compute uniform rate of profit
		r <- (1/(max(Mod(eigen(M)$values))))-1
		
		# See r
		r
		
		# Compute price of production vector
		D <- diag(3) - (1+r)*A
		p <- (1+r)*l%*%solve(D)
		
		# See price of production vector
		p
		
		# Compute vector of labor value
		B <- diag(3) - A
		v <- l %*% solve(B)
		
		# See value vector
		v
		
		# Compute value of real wage bundle
		c2 <- v%*%b
		
		# See value of real wage bundle
		c2
		
		# Definition of set B
		# Ensure that first condition for set B is satisfied
		# TRUE means the condition is satisfied
		ifelse(
		c2 < 1,TRUE,FALSE
		)
		
		# Ensure that second condition for set B is satisfied
		# TRUE means the condition is satisfied
		ifelse(
		1/c2 < max(p/v),TRUE,FALSE
		)
		
		# -------- CU-LS technical change in sector 3
		
		# New input-output vector
		A_new <- matrix(c(0.35,0.05,0.27,
		0.15, 0.45, 0.07,
		0.15, 0.15, 0.37), 
		byrow = TRUE, ncol = 3, nrow = 3)
		
		# See new input-output vector
		A_new
		
		# New labor input vector
		l_new <- matrix(c(0.2,0.15,0.18),ncol=3)
		
		# See new labor input vector
		l_new
		
		# Check that the viability condition is satisfied
		# If positive, viability condition is satisfied
		(p%*%A[,3] + l[1,3] - (p%*%A_new[,3] + l_new[1,3]))
		
		# Compute new vector of labor values
		B_new <- diag(3) - A_new
		v_new <- l_new %*% solve(B_new)
		
		# See new vector of labor values
		v_new
		
		# The constant needed to implement condition for 
		# cost reduction to be bounded above
		d1 <- (p%*%A[,3] + l[1,3] - p%*%A_new[,3])/l_new[1,3]
		
		# See d1
		d1
		
		# Coordinate of price hyperplane's intersection with axes
		B1 <- c(d1/p[1,1],d1/p[1,2],d1/p[1,3])
		
		# Coordinate of value hyperplane's intersection with axes
		B2 <- c(c2/v_new[1,1],c2/v_new[1,2],c2/v_new[1,3])
		
		# See B2
		B2
		
		# Condition (If max(B2/B1)>1, the problem is solved!)
		ifelse(
		max(B2/B1)>1,TRUE,FALSE
		)
		
		# ---- Example of a real wage bundle that keeps the
		# rate of exploitation constant but leads to a fall in the 
		# uniform rate of profit
		
		# Construct new real wage bundle (B1[2] < b2_new < B2[2])
		# We can draw from uniform distribution between B1 and B2
		set.seed(1000)
		(b2_new <- runif(1, min=B1[2], max=B2[2]))
		b1_new <- (c2 - b2_new*v_new[1,2])/(v_new[1,1]+v_new[1,3])
		b_exm <- matrix(c(b1_new,b2_new,b1_new),ncol = 1)
		
		# See new real wage bundle
		b_exm
		
		# New augmented input matrix
		M_new <- A_new + b_exm %*% l_new
		M_new
		
		# New uniform rate of profit
		r_new <- (1/(max(Mod(eigen(M_new)$values))))-1
		r_new
		
		# Compare old and new profit rates
		c(r,r_new)
		
		# Compare rate of exploitation
		c(v_new %*% b_exm, v%*%b)
		
		# Compute price of production vector
		D_new <- diag(3) - (1+r_new)*A_new
		p_new <- (1+r_new)*l_new%*%solve(D_new)
		
		# See price of production vector
		p_new
				
		# Ensure that second condition for set B is satisfied
		# TRUE means the condition is satisfied
		ifelse(
		1/(v_new %*% b_exm) < max(p_new/v_new),TRUE,FALSE
		)
		\end{verbatim}
		
	\end{appendices}
	
\end{document}